\documentclass[11pt, aps, prd, preprintnumbers, eqsecnum, nofootinbib]{revtex4-1}

\usepackage{epsf}
\usepackage{epsfig}
\usepackage{graphics}
\usepackage{graphicx}
\usepackage{amssymb}
\usepackage{mathtools}
\usepackage{color}
\usepackage{hyperref}
\usepackage{multirow}
\usepackage{feynmp-auto}
\usepackage[normalem]{ulem}

\usepackage{relsize}

\usepackage{amsmath, latexsym, amssymb, hyperref, graphicx, slashed, color, multirow}
\definecolor{nicered}{rgb}{0.0,.7,.3}
\definecolor{nicegreen}{rgb}{.1,.5,.1}
\definecolor{darkblue}{rgb}{0,.1,.9}
\hypersetup{colorlinks, citecolor=nicered ,linkcolor=darkblue, urlcolor=nicered}
\usepackage{cancel}

\def\lapp{\mathrel{\rlap{\raise.5ex\hbox{$<$}}
                    {\lower.5ex\hbox{$\sim$}}}}
\def\gapp{\mathrel{\rlap{\raise.5ex\hbox{$>$}}
                    {\lower.5ex\hbox{$\sim$}}}}

\newcommand{\beq}{\begin {equation}}  
\newcommand{\eeq}{\end   {equation}} 
\newcommand{\bea}{\begin {eqnarray}} 
\newcommand{\eea}{\end   {eqnarray}}  
\newcommand{\baa}{\begin {array}   } 
\newcommand{\eaa}{\end   {array}   }     
\newcommand{\bit}{\begin {itemize} }
\newcommand{\eit}{\end   {itemize} }
\newcommand{\be }{\begin {equation}} 
\newcommand{\ee }{\end   {equation}}
\newcommand{\nn }{\nonumber        }

\def\be{\begin{equation}}
\def\ee{\end{equation}}
\def\bea{\begin{eqnarray}}
\def\eea{\end{eqnarray}}

\def\beq{\begin{equation}}
\def\eeq{\end{equation}}
\newcommand{\beqa}{\begin{eqnarray}} 
\newcommand{\eeqa}{\end{eqnarray}}
\newcommand{\barr}{\begin{array}}
\newcommand{\earr}{\end{array}}

\def\gs{\mathrel{
   \rlap{\raise 0.511ex \hbox{$>$}}{\lower 0.511ex \hbox{$\sim$}}}}
\def\ls{\mathrel{
   \rlap{\raise 0.511ex \hbox{$<$}}{\lower 0.511ex \hbox{$\sim$}}}}

\newcommand{\braket}[1]{\left<#1\right>}

\begin{document}


\title{\Large Breaking flavor democracy with symmetric perturbations}


\author{\text{Tathagata Ghosh},$^{1,}$\footnote{Email: tghosh@hawaii.edu} \text{Jiajun Liao},$^{1,}$\footnote{Email: liaoj@hawaii.edu} \text{Danny Marfatia},$^{1,}$\footnote{Email: dmarf8@hawaii.edu} and  \text{Tsutomu T. Yanagida},$^{2,}$\footnote{Email: tsutomu.tyanagida@ipmu.jp}
}

\affiliation{$^1$Department of Physics \& Astronomy, University of Hawaii, Honolulu, HI 96822, USA
\\
$^2$Kavli IPMU (WPI), UTIAS, The University of Tokyo, Kashiwa, Chiba 277-8583, Japan
}


\begin{abstract}

Flavor democracy broken in the fermion mass matrix by means of small perturbations can give rise to hierarchical fermion masses. 
We study the breaking of the $\mathbb{S}^L_3 \times \mathbb{S}^R_3$ symmetry associated with democratic mass matrices to a smaller exchange symmetry $\mathbb{S}^L_2 \times \mathbb{S}^R_2$ in the charged lepton, up and down quark sectors. An additional breaking of the $\mathbb{S}^L_2 \times \mathbb{S}^R_2$ symmetry is necessary for the 
down quark mass matrix, which yields arbitrary perturbations in that sector. On the other hand, we require the neutrino mass matrix to be diagonal at the leading order, with the perturbations left arbitrary due to the absence of any guiding symmetry.
We show that the interplay between these two kinds of perturbations reproduces the quark and lepton mass and mixing observables for either hierarchy of neutrino masses.


\end{abstract}


\maketitle

\newpage
\section{Introduction}
\label{sec:intro}

Flavor mixing has long been established in both quark and lepton sectors. However, the standard model (SM) of particle physics does not offer an explanation of the observed flavor structure in either sector.
In fact, the Yukawa couplings of the recently discovered Higgs boson are the least understood part of the SM. Over the years various attempts have been made to explain the flavor structure of the SM fermions. In principle, one can follow either a top-down or a bottom-up~\cite{PMM,Fukugita:1998vn,Tanimoto} approach. In the former case, one starts by imposing a flavor symmetry to the theory, followed by prescriptions to break that symmetry to generate the Cabibbo-Kobayashi-Maskawa (CKM) and Pontecorvo-Maki-Nakagawa-Sakata (PMNS) matrices. The observed mixing patterns are then explained in terms of some residual symmetry. In contrast, in the latter case one attempts to reconstruct residual symmetries from the mixing patterns in both up-type and down-type fermion sectors~\cite{Lam}, and then the full flavor symmetry is obtained as a product group of residual symmetries~\cite{Smirnov}. This is called the \emph{phenomenological mass matrix approach}. Residual symmetries, however, control the mixing pattern in both cases~\cite{Dicus:2010yu} and can establish a sum rule~\cite{Sum Rule} between the Dirac CP phase and mixing angles. We adopt the second approach in this paper.

It is well known that the mixing pattern in the lepton and quark sector differ significantly. While the lepton sector exhibits large mixing angles, the mixing angles in the quark sector are small. The observed lepton and quark mixings are a combination of mixings in the down and up-type fermions, with $V_{\text{PMNS}}$ and $V_{\text{CKM}}$ in the lepton and quark sectors given by $V^{\dagger}_{\ell} V_{\nu}$ and $V^{\dagger}_u V_d$, respectively. Although the disparity between the above mixing patterns may appear puzzling, a unified framework can be constructed if both up- and down-type quark matrices and one of the lepton or neutrino mass matrices possess the same mixing pattern. Then large mixing angles can be generated in $V_{\text{PMNS}}$, and at the same time keeping $V_{\text{CKM}}$ close to the identity.   

Approximate democratic mass matrices~\cite{demo,Koide:1989ds,Tanimoto:1989qh,plankl,
Fritzsch:1995dj,Xing:1996hi,Mondragon:1998gy,
Fritzsch:1998xs,Fritzsch:1999ee} for both up- and down-type quarks offer an exciting avenue to explain the small CKM mixing angles and the large hierarchy of quark masses. The application of the same hypothesis to both charged lepton and neutrino mass matrices leads to too small mixing angles in $V_{\text{PMNS}}$. Instead, if the neutrino mass matrix assumes an almost diagonal form, then $V_{\text{PMNS}}$ can easily have two large mixing angles~\cite{Fukugita:1998vn,Tanimoto}. The charged lepton mass matrix remains democratic, leading to a natural explanation of the hierarchy in their masses, with $m_1=m_2 =0$ at the leading order. However, one needs to break the residual $\mathbb{S}_3$ symmetries in the up, down, and charged lepton mass matrices to accommodate nonzero light fermion masses and to fit the experimentally measured mixing angles. 

Such a study is performed in Ref.~\cite{Ge} under the assumption that there is no remnant symmetry in the up, down, and charged lepton mass matrices, after the $\mathbb{S}_3^L \times \mathbb{S}_3^R$ symmetry in the democratic matrices is broken. Hence, there is no guiding symmetry to regulate the small perturbations, and the perturbations are random.  
In contrast, we consider a residual $\mathbb{S}_2^L \times \mathbb{S}_2^R$ symmetry in the up, down and charged lepton mass matrices~\cite{plankl}. However, the $\mathbb{S}_2^L \times \mathbb{S}_2^R$ symmetry in the down sector needs to be broken eventually. 
Also, we assume that there is no residual symmetry left in the neutrino sector allowing the perturbations in that sector to be arbitrary. We take neutrinos to be Majorana particles.  
 (Previously, Ref.~\cite{Fritzsch:1999ee} studied $\mathbb{S}_3^L \times \mathbb{S}_3^R \rightarrow \mathbb{S}_2^L \times \mathbb{S}_2^R $ breaking with different subsequent breaking patterns than we consider.)
Our approach is different from the anarchy scenario~\cite{Anarchy} in which the mass matrix elements are unconstrained.



The paper is organized as follows. In Section~\ref{sec:DemMass} we review the origin and predictions of democratic mass matrices. We discuss the consequences of breaking the $\mathbb{S}_3^L \times \mathbb{S}_3^R$ to a smaller $\mathbb{S}_2^L \times \mathbb{S}_2^R$ symmetry in Section~\ref{sec:perturb}, and show that solutions can be found that are consistent with all mass and mixing constraints in the lepton and quark sectors. We conclude in Section~\ref{sec:conclusion}.
  

\section{Democratic Mass Matrix }
\label{sec:DemMass}

Without Yukawa matrices the SM has an $U(3)^5$ global symmetry. One can start building a model with flavor structure by using a maximal subgroup of $U(3)^5$. Following Ref.~\cite{Tanimoto}, our starting point is $O(3)_{L\{Q,L\}} \times O(3)_{R\{u^c,d^c,e^c\}}$. In the above model, the $O(3)_L$ symmetry in the neutrino sector is explicitly broken by $\Sigma_L$, which transforms as $(\bf{5,1})$ under the symmetry groups and takes the form diag$\{a,b,-a-b\}$. This gives a diagonal neutrino mass matrix at the leading order. Besides, the explicit breaking of the flavor symmetry prevents the existence of unwanted Nambu-Goldstone bosons. On the other hand, the charged leptons obtain mass at the leading order from another explicit breaking of $O(3)_{L} \times O(3)_{R}$ by $\phi_{L,R}$, which transform as $(\bf{3,1})$ and $(\bf{1,3})$, respectively. The vacuum expectation values (VEV) of $\phi_{L,R}$ have a structure $\braket{\phi_{L,R}}= (1,1,1)^T \, v_{L,R}$, resulting in a democratic mass matrix for the charged leptons preserving an $\mathbb{S}_3^L \times \mathbb{S}_3^R$ symmetry. The up and down-type quarks obtain mass at the leading order similarly to charged leptons. 

The democratic fermion mass matrix, possessing $\mathbb{S}_3^L \times \mathbb{S}_3^R$ symmetry has the form~\cite{demo,Koide:1989ds,Tanimoto:1989qh,plankl,
Fritzsch:1995dj,Xing:1996hi,Mondragon:1998gy,
Fritzsch:1998xs,Fritzsch:1999ee},
\beq
M_f = \dfrac{M_{0}}{3} \, \begin{pmatrix}
1 & 1 & 1\\
1 & 1 & 1\\
1 & 1 & 1
\end{pmatrix} \, ,
\label{eq:demMM}
\eeq
where $M_{0}$ is the characteristic mass scale~\cite{Fukugita:1998vn,Tanimoto}. This matrix can be diagonalized by $M_f = V_L D_f V^{\dagger}_R$, where $V_L \, (V_R)$ are mixing matrices for left (right) handed fermions, and $D_f = \text{diag}\{m_1,m_2,m_3\}$ is the diagonalized mass matrix. Now, $V_L$ can be determined by diagonalizing $M_f M^{\dagger}_f = V_L D^2_f V^{\dagger}_L$, where $M_f M^{\dagger}_f$ has the same form as $M_f$ with a normalization constant $M^2_0/3$. Thus, by diagonalizing $M_f M^{\dagger}_f$ the most general form of $V_L$ is 
\bea
V^{\dagger}_L & = & \begin{pmatrix}
e^{i \beta_1} & & \\
& e^{i \beta_2}& \\
& & e^{i \beta_3}
\end{pmatrix} \,
\,   \begin{pmatrix}
-\dfrac{1}{\sqrt{2}} & \dfrac{1}{\sqrt{2}} & 0\\
-\dfrac{1}{\sqrt{6}} & -\dfrac{1}{\sqrt{6}} & \dfrac{2}{\sqrt{6}}\\
 \dfrac{1}{\sqrt{3}}& \dfrac{1}{\sqrt{3}} & \dfrac{1}{\sqrt{3}}
\end{pmatrix} \, \equiv \, R  V_0 \, ,
\label{eq:VL}
\eea
where 
each mass eigenvalue is associated with a rephasing degree of freedom denoted by $\beta_i \, (i=1,2,3)$. It is obvious from Eqs.~(\ref{eq:demMM}) and (\ref{eq:VL}) that the squared eigenvalues generated by diagonalizing $M_f M^{\dagger}_f$ are $\{0,0, M^2_0\}$, thus producing a natural hierarchy in the respective fermion sectors.

In the lepton sector, since $V_{\nu} = I$ at the leading order,  Eq.~(\ref{eq:VL}) is the PMNS matrix. The mixing angles in terms of the standard PMNS parametrization are~\cite{Fritzsch:1995dj,Fritzsch:1998xs}
\beq
\theta_{13}= 0^\circ \,, \ \ \ \theta_{12} = 45^\circ \,, \ \ \ \theta_{23} = 54.7^\circ   \,,  
\eeq
which are clearly in tension with the measured values of the mixing angles for neutrinos~\cite{deSalas:2017kay}. Hence, it is imperative to break the democracy in the charged lepton mass matrix of Eq.~(\ref{eq:demMM}). This is also needed to provide a nonzero mass to $m_{\mu}$.

On the other hand, in the quark sector both up- and down-type quark mass matrices have a democratic form, so that $V_{\text{CKM}} $ is simply the identity matrix with no rephasing matrices. Also, $m_{1,2} = 0$ in both quark sectors. Again, we need a deviation from democracy in the quark sector to explain the observed quark masses and mixing angles. 
\section{Broken democracy }
\label{sec:perturb}

In this section, we discuss the consequences of breaking the democracy in the mass matrices with $\mathbb{S}_2^L \times \mathbb{S}_2^R$ invariant perturbations, and comment on perturbations  preserving other smaller discrete exchange symmetries.


It is reasonable to assume that once the $\mathbb{S}_3^L \times \mathbb{S}_3^R$ symmetry in the democratic mass matrix is broken, there will not be any remnant symmetry in the charged lepton mass matrix. In Ref.~\cite{Ge}, anarchic random perturbations are employed to write $M_{f_{ij}} = \dfrac{M_0}{3} (1+\delta_{ij})$, where $0 \leq |\delta_{ij}| \leq \delta_{max}$ and the phases of $\delta_{ij}$ are randomly scanned. The peak of the mixing angle and Dirac CP phase distributions are found to be stable against different values of $\delta_{max}$.

In contrast, we assume that the $\mathbb{S}_3^L \times \mathbb{S}_3^R$ symmetry in the democratic mass matrix breaks to a smaller exchange symmetry $\mathbb{S}_2^L \times \mathbb{S}_2^R$. 
Thus, the perturbations to the charged lepton and quark mass matrices are $\mathbb{S}_2^L \times \mathbb{S}_2^R $ invariant in our model.
 Although $m_e=0$ in this case, the measured electron mass can be easily generated by radiative corrections. 
 The general form of charged lepton, up, and down quark mass matrices with $\mathbb{S}_2^L \times \mathbb{S}_2^R$ perturbations is given by
\beq
M_f = \dfrac{M_{0}}{3} \, \Bigg [ \begin{pmatrix}
1 & 1 & 1\\
1 & 1 & 1\\
1 & 1 & 1
\end{pmatrix} \, + \,   \begin{pmatrix}
a & a & b\\
a & a & b\\
c & c & d
\end{pmatrix} \Bigg] \, ,
\label{eq:demS2xS2MM}
\eeq
where $\{a,b,c,d\} \in \mathbb{C}$. As usual, to obtain the PMNS matrix we need to diagonalize $M_f M^{\dagger}_f$. We do this sequentially to get some insight about the mass matrix. First, we reduce $M_f M^{\dagger}_f $ to the form,
\beq
\tilde{M}_f^2 \, =  \dfrac{M^2_0}{9}\, \begin{pmatrix}
0 & 0 & 0\\
0 & X & Y \\
0 & Y^* & 9+Z
\end{pmatrix} \, ,
\label{eq:demS2xS2MM_2}
\eeq 
by evaluating $V_0 M_f M^{\dagger}_f V^{\dagger}_0 $, where $V_0$ is defined in Eq.~(\ref{eq:VL}). Here, $X, Y$ and $Z$ are
\bea
X &\equiv& \dfrac{2}{3} \big [ |b|^2 + |d|^2 + 2 |a-c|^2 - 2 Re(bd^*)\big ]\,, \nn \\
Y & \equiv & \dfrac{\sqrt{2}}{3} \big [ |d|^2 +2 |c|^2 -4 |a|^2 - 2 |b|^2 - 3(2a+b-2c-d) \nn \\
& & + 4 a^*c -2 ac^* + 2b^*d -bd^*\big ]\,, \nn \\
Z & \equiv & \dfrac{2}{3}|2a+c|^2 + \dfrac{1}{3} |2b+d|^2 + 2 Re(4a+2b+2c+d) \, .
\label{eq:XYZ}
\eea 
From Eq.~(\ref{eq:demS2xS2MM_2}) it is obvious that $m_e=0$ with the remnant $\mathbb{S}_2^L \times \mathbb{S}_2^R$ symmetry. Now, $\tilde{M}_f^2$ can be diagonalized by $\tilde{M}_f^2= T^{\dagger} D_f^2 T$, where
\beq
T = \begin{pmatrix}
1 & 0 & 0 \\
0 & c_{\theta_{T}} & s_{\theta_{T}} e^{i \alpha_{T}} \\
0 & -s_{\theta_{T}} e^{-i \alpha_{T}} & c_{\theta_{T}}
\end{pmatrix} \, .
\eeq
We obtain the following masses for the charged leptons, up and down quarks:
\bea
m^2_1 & = & 0 \, , \nn \\
m^2_{2} & = &  \dfrac{M^2_0}{9} \Bigg[ \dfrac{9+Z+X}{2} - \sqrt{\bigg(\dfrac{9+Z-X}{2}\bigg)^2+|Y|^2} \Bigg]  \, , \nn \\
m^2_{3} & = &   \dfrac{M^2_0}{9} \Bigg[ \dfrac{9+Z+X}{2} + \sqrt{\bigg(\dfrac{9+Z-X}{2}\bigg)^2+|Y|^2} \Bigg] \, .
\label{eq:fermion_mass_S2xS2}
\eea 
The PMNS matrix becomes
\bea
V_{\text{PMNS}} & = & \begin{pmatrix}
-\dfrac{1}{\sqrt{2}}  & \dfrac{1}{\sqrt{2}}  &  0   \\
-\dfrac{c_{\theta_{T,\ell}}}{\sqrt{6}} + \dfrac{s_{\theta_{T,\ell}} e^{i \alpha_{T,\ell}}}{\sqrt{3}} & -\dfrac{c_{\theta_{T,\ell}}}{\sqrt{6}} + \dfrac{s_{\theta_{T,\ell}} e^{i \alpha_{T,\ell}}}{\sqrt{3}} & \dfrac{2 c_{\theta_{T,\ell}}}{\sqrt{6}} + \dfrac{s_{\theta_{T,\ell}} e^{i \alpha_{T,\ell}}}{\sqrt{3}} \\ 
\dfrac{c_{\theta_{T,\ell}}}{\sqrt{3}} + \dfrac{s_{\theta_{T,\ell}} e^{-i \alpha_{T,\ell}}}{\sqrt{6}} & \dfrac{c_{\theta_{T,\ell}}}{\sqrt{3}} + \dfrac{s_{\theta_{T,\ell}} e^{-i \alpha_{T,\ell}}}{\sqrt{6}} & \dfrac{c_{\theta_{T,\ell}}}{\sqrt{3}} - \dfrac{2 s_{\theta_{T,\ell}} e^{-i \alpha_{T,\ell}}}{\sqrt{6}}
\end{pmatrix} \, . 
\eea
It is clear from the above matrix that $\theta_{12} = 45^\circ$, $\theta_{13} = 0^\circ$, and $\theta_{23}$ is given by
\beq
\cos \theta_{23} = \bigg| \dfrac{c_{\theta_{T,\ell}}}{\sqrt{3}} - \dfrac{2s_{\theta_{T,\ell}} e^{-i \alpha_{T,\ell}}}{\sqrt{6}} \bigg| \, .
\eeq 
Similarly, the only nonzero mixing angle in the CKM matrix is given by
\beq
\cos \theta_{23} = \big| c_{\theta_{T,u}}c_{\theta_{T,d}}+s_{\theta_{T,u}}s_{\theta_{T,d}} e^{i(\alpha_{T,u}-\alpha_{T,d})} \big| \, .
\eeq 

It is evident from the discussion above that $ \mathbb{S}_2^L \times \mathbb{S}_2^R$ perturbations to democratic mass matrices themselves cannot explain the observed masses and mixings of the SM fermions and that additional perturbations need to be included to account for the experimental values of the fermion masses and mixings. 
We find that each of $a,b,c$ and $d$ alone of magnitude $0.2- 0.3$ can fit the mass ratio $m_{\mu}/m_{\tau}$ in the $3 \sigma$ range. However, the relations of Eq.~(\ref{eq:fermion_mass_S2xS2}) do not reveal any correlation between $a,b,c$ and $d$. For simplicity we focus on two special cases of the $ \mathbb{S}_2^L \times \mathbb{S}_2^R$ perturbation matrix in Eq.~(\ref{eq:demS2xS2MM}):
\begin{itemize}
\item \emph{Case~I:} We examine scenarios where only one of $a,b,c$ and $d$ is nonzero. We find that the results from all such scenarios are  similar. Hence, we discuss only the case with $a=b=c=0$, and $d \neq 0$, which leads to a fermion mass matrix of the form,
\beq
M_f = \dfrac{M_{0}}{3} \, \Bigg [ \begin{pmatrix}
1 & 1 & 1\\
1 & 1 & 1\\
1 & 1 & 1
\end{pmatrix} \, + \,   \begin{pmatrix}
0 & 0 & 0\\
0 & 0 & 0\\
0 & 0 & |d| e^{i \alpha_{I}}
\end{pmatrix} \Bigg] \, ;
\label{eq:demS2xS2MMCaseI}
\eeq
\item \emph{Case~II:} We consider a special case where the democratic $ \mathbb{S}_3^L \times \mathbb{S}_3^R$ symmetry in the charged lepton, up, and down quark mass matrices are broken by fields $\phi'_{L,R}$, with $\braket{\phi'_{L,R}} = (a,a,b)^T $. $\phi'_{L,R}$ are triplets of $O_3^L \times O_3^R$ and so $a,b \in \mathbb{R}$.
Consequently the fermion mass matrix reduces to the form,
\beq
M_f = \dfrac{M_{0}}{3} \, \Bigg [ \begin{pmatrix}
1 & 1 & 1\\
1 & 1 & 1\\
1 & 1 & 1
\end{pmatrix} \, + \,   e^{i \alpha_{II}} \begin{pmatrix}
a^2 & a^2 & ab\\
a^2 & a^2 & ab\\
ab & ab & b^2
\end{pmatrix} \Bigg] \, ,
\label{eq:demS2xS2MMCaseII}
\eeq
where $\alpha_{II}$ is the relative phase between the first and second terms. Note that mass terms for fermions can also be generated from the interactions $\phi'_L \otimes \phi_R$ and $\phi_L \otimes \phi'_R$, but such terms are removed by imposing the parity symmetry.
\end{itemize}


We now show that by including arbitrary small perturbations to the neutrino and down quark mass matrices, along with $ \mathbb{S}_2^L \times \mathbb{S}_2^R$ perturbations in the charged lepton and up quark mass matrices, all the mass and mixing observables in both the lepton and quark sectors can be reproduced. It is not unnatural to assume arbitrary perturbations in the neutrino mass matrix since there is no guiding symmetry left in the neutrino sector. On the other hand, both up and down quarks obtain their masses in a manner similar to charged leptons. So the perturbations in both up and down mass matrices are expected to preserve the $\mathbb{S}_2^L \times \mathbb{S}_2^R$ symmetry. However, this yields $\theta_{12}=\theta_{13}=0$ in the CKM matrix. Consequently, to fit the CKM matrix, we are forced to assume that the $\mathbb{S}_2^L \times \mathbb{S}_2^R$ symmetry in the democratic down quark mass matrix is entirely broken, and the resulting perturbations are a combination of $\mathbb{S}_2^L \times \mathbb{S}_2^R$ invariant and arbitrary perturbations. In principle, the $\mathbb{S}_2^L \times \mathbb{S}_2^R$ symmetry in the charged lepton and up quark sectors can also be broken, but such a breaking must be much softer than in the down quark sector since the hierarchy in the down quark sector is milder than the other two sectors.


\subsection{Mass and mixing observables in the lepton sector}
\label{sec:lepton}


{
We numerically compute the eigenvalues of the charged lepton mass matrices of Eqs.~(\ref{eq:demS2xS2MMCaseI}) and (\ref{eq:demS2xS2MMCaseII}) using  Eqs.~(\ref{eq:XYZ}) and (\ref{eq:fermion_mass_S2xS2}). 
We compare the predicted value of $m_{\mu}/m_{\tau}$ in our model with the renormalized value $m_{\mu}/m_{\tau}$ evaluated at a common energy scale $M_Z$ using the experimentally measured values of $m_{\mu}$ and $m_{\tau}$, and require that the predicted value lies within the $3 \sigma$ range of the associated uncertainty~\cite{Running_masses}. 
This constraint produces a correlation between the $ \mathbb{S}_2^L \times \mathbb{S}_2^R$ perturbation parameters. 
In Fig.~\ref{fig:correlation-1}, we show the correlations between ($|d_\ell|,\alpha_{I_\ell}$) and ($a_\ell,b_\ell$) for \emph{Cases-I} and \emph{II}, respectively. 
\begin{figure}[t]
\includegraphics[scale=0.5]{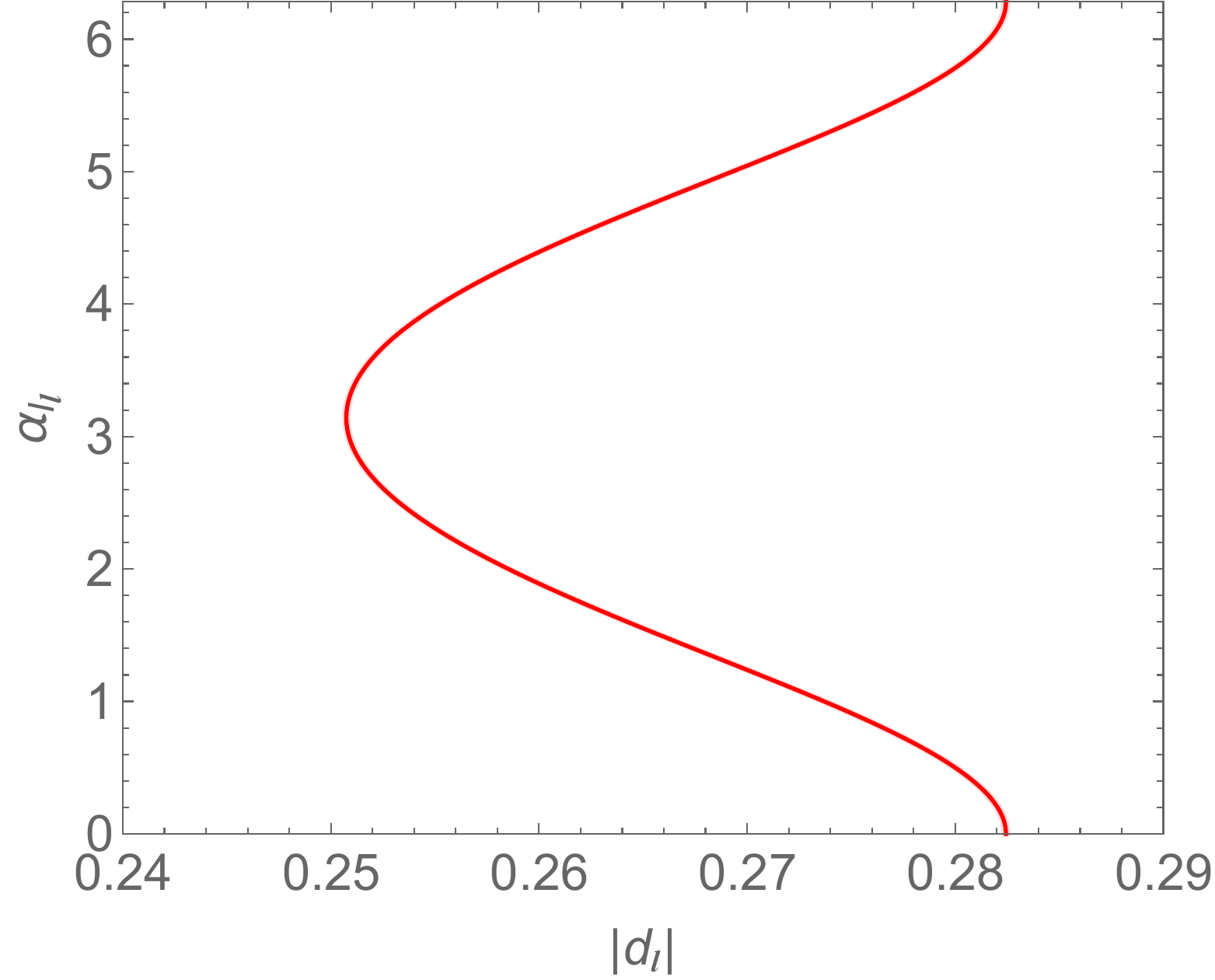}
\includegraphics[scale=0.5]{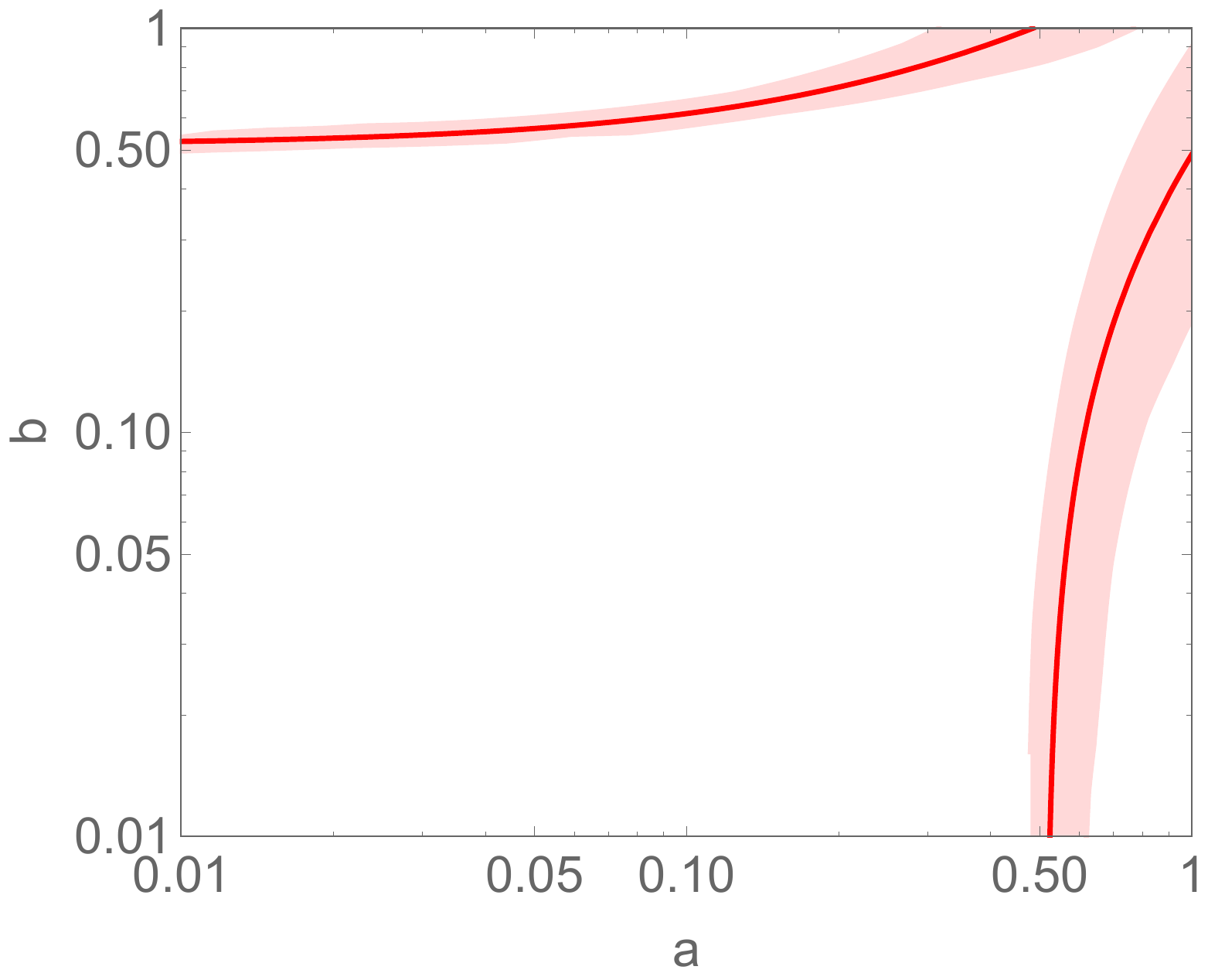}
\caption{The correlation between the $ \mathbb{S}_2^L \times \mathbb{S}_2^R$ perturbation parameters in the charged lepton mass matrix with $m_{\mu}/m_{\tau}$  within its $3 \sigma$ measured range. The left panel shows the correlation between $|d_\ell|$ and $\alpha_{I_\ell}$ for \emph{Case~I}, and the right panel depicts the correlation between $a_\ell$ and $b_\ell$ for \emph{Case~II}. We do not find any correlation between $a_\ell$ or $b_\ell$ with $\alpha_{{II}_\ell}$
 at the leading order.
}
\label{fig:correlation-1}
\end{figure}
The central curve in both plots of Fig.~\ref{fig:correlation-1} can be explained by the leading order expansions,
\begin{align}
\frac{m_{\mu}}{m_{\tau}}=\frac{2}{9}|d_{\ell}|\left(1-\frac{2}{9}|d_{\ell}|\cos\alpha_{I_{\ell}}\right)+{\cal O} (|d_{\ell}|^3) \, ,
\label{eq:M2overM3-I}
\end{align}
for \textit{Case~I}, and
\begin{align}
\frac{m_{\mu}}{m_{\tau}}=\frac{2}{9}\left(a_{\ell}-b_{\ell}\right)^2+{\cal O}(a_{\ell}^3,b_{\ell}^3,a_{\ell}^2b_{\ell},a_{\ell}b_{\ell}^2) \, ,
\label{eq:M2overM3-II}
\end{align}
for \textit{Case~II}. 
}
We conclude from Fig.~\ref{fig:correlation-1} that for \emph{Case~I}, $0.25 \lesssim |d_\ell| \lesssim 0.29$ is required with a weak dependence on $\alpha_{I_\ell}$. For \emph{Case~II}, a relatively broad range of values are allowed for $a_\ell$ and $b_\ell$, which is expected owing to the larger number of parameters. From Eq.~(\ref{eq:M2overM3-II}) we see that there is no phase dependence at the leading order for this case. However, when higher order terms are included in the numerical calculation, we observe that
$\alpha_{{II}_\ell}$ is constrained within $\pi/2 \lesssim \alpha_{{II}_\ell} \lesssim 3 \pi/2$, for $0.3 < a_\ell,b_\ell < 0.5$. For all other values of $(a_\ell,b_\ell)$,  $\alpha_{{II}_\ell}$ remains unconstrained.

For the neutrino mass matrix we consider both the normal hierarchy (NH) and the inverted hierarchy (IH). The initial neutrino mass matrix is 
\beqa
 M_{\nu}^{(0)} &=&  \text{diag} \bigg( m_1^0 , \sqrt{(m_1^0)^2 + \overline{\Delta m^2_{21}}}, \sqrt{(m_1^0)^2 + \overline{\Delta m^2_{31}}} \bigg) \, \ \ \ \ \ \ \    \rm{for\ NH}\,\\
 &=& \text{diag} \bigg( \sqrt{(m_3^0)^2 +\overline{\Delta m^2_{31}}}, \sqrt{(m_3^0)^2 + \overline{\Delta m^2_{31}} + \overline{\Delta m^2_{21}}}, m_3^0 \bigg) \ \ \ \ \ \ \ \rm{for\ IH}\,, 
 \eeqa
 where $\overline{\Delta m^2_{21}} \,\ (\overline{\Delta m^2_{31}})$ represents the best-fit value  of the mass squared difference, ${\Delta m^2_{21}}\, ({\Delta m^2_{31}})$. Here, $m_1^0\ (m_3^0)$ is the lightest neutrino mass for NH (IH). 
Then we add arbitrary perturbations to the complex symmetric neutrino mass matrix $M_{\nu_{ij}}^{(1)} = \epsilon_{\nu_{ij}} e^{i \alpha_{\nu_{ij}}}$, where $\epsilon_{\nu_{ij}} \leq  0.01$ eV and $\alpha_{\nu_{ij}} \in [0, 2\pi]$. 
The full neutrino mass matrix is
\beq
M_{\nu} = M_{\nu}^{(0)} + M_{\nu}^{(1)} \, .
\eeq

After adding arbitrary perturbations to the neutrino mass matrix, we diagonalize it and impose cuts to ensure that  ${\Delta m^2_{21}}$ and ${\Delta m^2_{31}}$  are within the $3 \sigma$ ranges of their experimental values~\cite{deSalas:2017kay}. Then, we impose additional cuts to select perturbation parameters that also give $\theta_{12}$, $\theta_{13}$, and $\theta_{23}$ within $3 \sigma$ ranges of their measured values~\cite{deSalas:2017kay}. 
We find a few solutions that satisfy all the above criteria, but we do not find any preference for $\delta_{\text{CP}}$ or the Majorana phases $\varphi_1$ and $\varphi_2$.
In Table~\ref{tab:lepton} we list the perturbation parameters of benchmark points for each case and each hierarchy that satisfy all the above constraints. We use the {\tt MPT} package to compute the Majorana phases for our benchmark points~\cite{Antusch:2005gp}.

\begin{table}[t]
\resizebox{\columnwidth}{!}{%
\begin{tabular}{|c |c|c| c| c|c|c c c|c c c|c| c c|c c c c c c| }
\hline
Case & Hierarchy &  \multicolumn{3}{|c|}{Charged lepton perturbations} & & \multicolumn{6}{|c|}{Neutrino sector perturbations}  & & \multicolumn{2}{|c|}{} & \multicolumn{6}{|c|}{$V_{\text{PMNS}}$ observables}
\\
\hline
 &  & $|d_{\ell}|$ or $a_{\ell}$ & $b_{\ell}$ & $\alpha_{I_{\ell}}$ or $\alpha_{II_{\ell}}$& $m_{\nu}^{\text{lightest}}$ & \multicolumn{3}{|c|}{$\epsilon_{\nu_{ij}}$}  & \multicolumn{3}{|c|}{$\alpha_{\nu_{ij}}$} & $\dfrac{m_{\mu}}{m_{\tau}}$ & $\Delta m^2_{21}$ &  $|\Delta m^2_{31}|$ & $\theta_{12}$ & $\theta_{13}$ & $\theta_{23}$ & $\delta_{\text{CP}}$ & $\varphi_1$ & $\varphi_2$ \\
 & & & & [Radian] & [eV] & \multicolumn{3}{|c|}{[eV]} & \multicolumn{3}{|c|}{[Radian]} & & \multicolumn{2}{|c|}{[eV$^2$]} & \multicolumn{6}{|c|}{$[\circ]$} \\
\hline 
\hline

\multirow{6}{*}{\emph{I}} & \multirow{3}{*}{NH} & \multirow{3}{*}{0.270} & \multirow{3}{*}{-} & \multirow{3}{*}{4.98} & \multirow{3}{*}{$5.15 \times 10^{-2}$}  & 0.0095 & 0.0004 & 0.00004 & 1.35 & 5.14 & 0.89 & \multirow{3}{*}{0.058825} & \multirow{3}{*}{$7.28 \times 10^{-5}$} & \multirow{3}{*}{$2.47 \times 10^{-3}$} & \multirow{3}{*}{34.1} & \multirow{3}{*}{8.20}  & \multirow{3}{*}{45.3} & \multirow{3}{*}{8.67} & \multirow{3}{*}{15.9} & \multirow{3}{*}{1.66} \\

 & & &  & & & -- & 0.0039 & 0.0039 & -- & 5.31 & 0.04 & & & & & & & & &\\
 & &  &  & & & -- & -- & 0.0005 & -- & -- & 5.18 & & & & & & & & &\\
\cline{2-21}

 & \multirow{3}{*}{IH} & \multirow{3}{*}{0.275} & \multirow{3}{*}{-} & \multirow{3}{*}{0.99} & \multirow{3}{*}{$1.97 \times 10^{-5}$} & 0.0022 & 0.0074 & 0.0054 & 2.22 & 4.87 & 3.96 & \multirow{3}{*}{0.058816} & \multirow{3}{*}{$7.08 \times 10^{-5}$} & \multirow{3}{*}{$2.44 \times 10^{-3}$} & \multirow{3}{*}{32.6} & \multirow{3}{*}{8.31}  & \multirow{3}{*}{52.9} & \multirow{3}{*}{37.7} & \multirow{3}{*}{308.8} & \multirow{3}{*}{305.9} \\

 & & &  & & & -- & 0.0044 & 0.0053 & -- & 1.93 & 0.91 & & & & & & & & &\\
 & &  &  & & & -- & -- & 0.0045 & -- & -- & 5.59 & & & & & & & & &\\
\hline

\multirow{6}{*}{\emph{II}} & \multirow{3}{*}{NH} & \multirow{3}{*}{0.489} & \multirow{3}{*}{0.022} & \multirow{3}{*}{2.67} &  \multirow{3}{*}{$1.31 \times 10^{-2}$} & 0.0016 & 0.0016 & 0.0095 & 1.61 & 1.30 & 6.06 & \multirow{3}{*}{0.058833} & \multirow{3}{*}{$7.25 \times 10^{-5}$} & \multirow{3}{*}{$2.60 \times 10^{-3}$} & \multirow{3}{*}{37.3} & \multirow{3}{*}{8.47}  & \multirow{3}{*}{41.3} & \multirow{3}{*}{2.42} & \multirow{3}{*}{352.7} & \multirow{3}{*}{9.50}\\

 & & &  & & & -- & 0.0021 & 0.0015 & -- & 3.68 & 5.22 & & & & & & & & &\\
 & & &  & & & -- & -- & 0.0067 & -- & -- & 1.97 & & & & & & & & &\\
 
\cline{2-21}

 & \multirow{3}{*}{IH} & \multirow{3}{*}{0.537} & \multirow{3}{*}{0.096} & \multirow{3}{*}{3.20} & \multirow{3}{*}{$4.18 \times 10^{-4}$} & 0.0018 & 0.0008 & 0.0049 & 5.13 & 5.61 & 3.00 & \multirow{3}{*}{0.058836} & \multirow{3}{*}{$7.19 \times 10^{-5}$} & \multirow{3}{*}{$2.54 \times 10^{-3}$} & \multirow{3}{*}{32.6} & \multirow{3}{*}{7.91}  & \multirow{3}{*}{49.2} & \multirow{3}{*}{160.6} & \multirow{3}{*}{56.2} & \multirow{3}{*}{63.1}\\

 & & &  & & & -- & 0.0035 & 0.0052 & -- & 1.81 & 5.77 & & & & & & & & &\\
 & & &  & & & -- & -- & 0.0028 & -- & -- & 0.92 & & & & & & & & &\\
\hline
\end{tabular}
}
\caption{ An illustrative set of perturbation parameters for charged lepton and neutrino sectors, and the corresponding values of neutrino mass and mixing observables for \emph{Cases I} and \emph{II}. 
{ Since the neutrino mass matrix is complex symmetric, we only show the perturbation parameters in the neutrino sector for $i \leq j$. }  }
\label{tab:lepton}
\end{table}





\subsection{Mass and mixing observables in the quark sector}
\label{sec:quark}

\begin{figure}[t]
\includegraphics[scale=0.5]{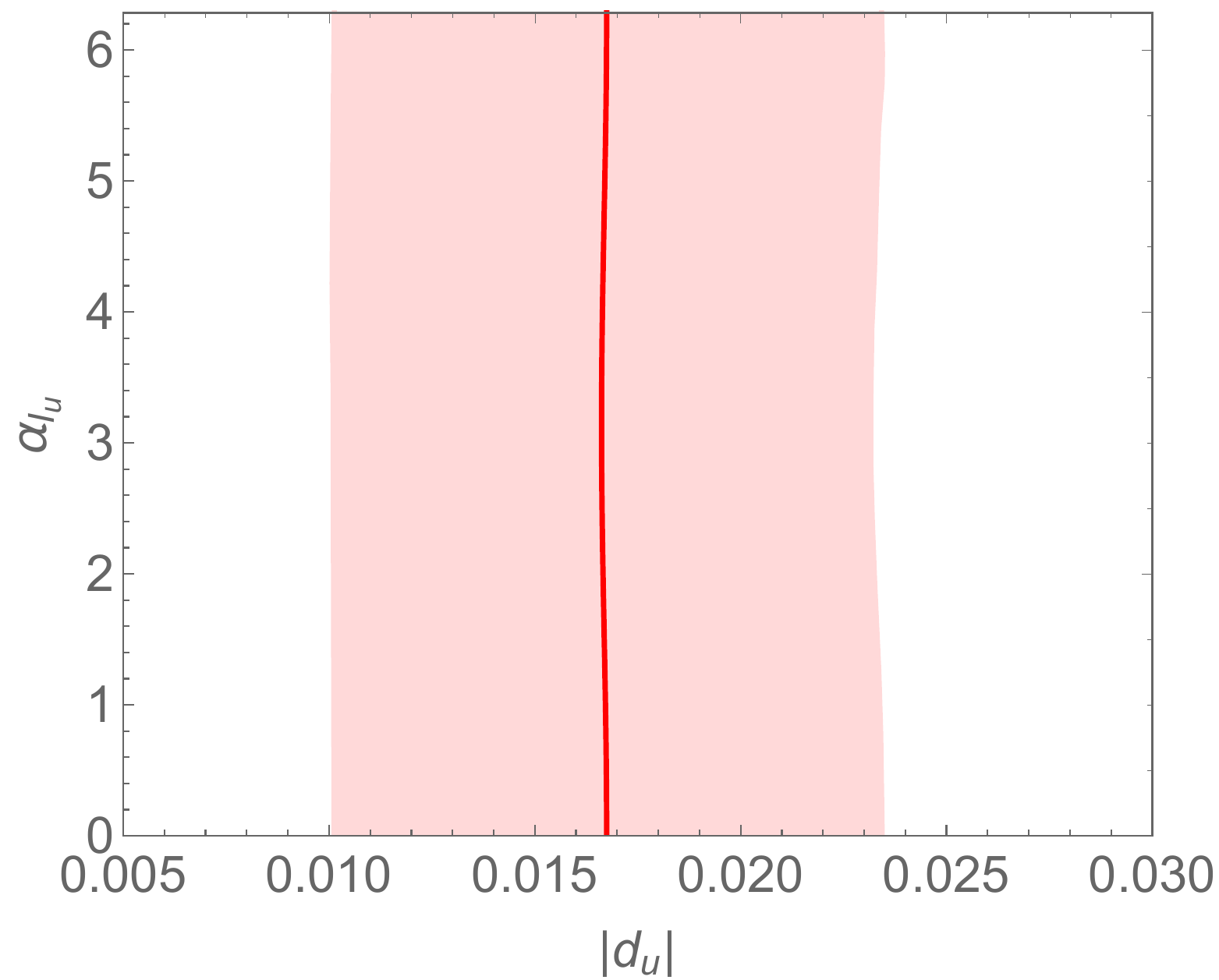}
\includegraphics[scale=0.5]{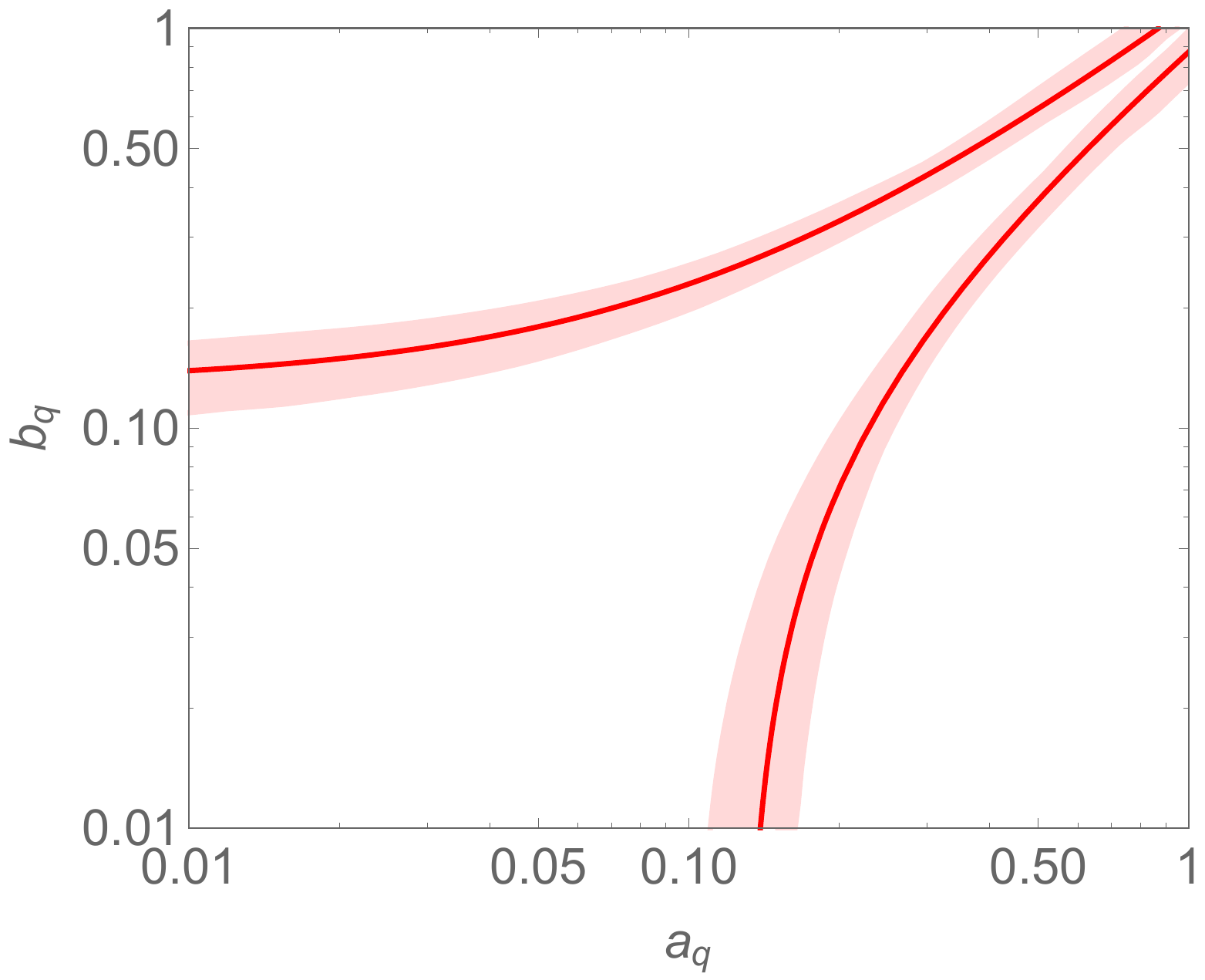}
\caption{The same as Fig.~\ref{fig:correlation-1}, but with $m_{c}/m_{t}$ within its $3 \sigma$ measured range.}
\label{fig:correlation_quark-1}
\end{figure}

Now, we discuss our fit to the quark mass and mixing observables. For the up quark mass matrix, we follow the same 
procedure as used for the charged lepton mass matrix in the previous subsection. Also, we obtain relations identical to Eqs.~(\ref{eq:M2overM3-I}) and (\ref{eq:M2overM3-II}) with $m_{\mu}/m_{\tau}$ replaced by $m_c/m_t$, and the charged lepton perturbation parameters replaced by the corresponding up quark ones. 
Then we apply the constraint that the predicted $m_{c}/m_{t}$  is satisfied within the $3 \sigma$ range of the renormalized value of $m_{c}/m_{t}$ at $M_Z$ computed from the experimentally measured value~\cite{Running_masses}.
The resultant correlation between the $ \mathbb{S}_2^L \times \mathbb{S}_2^R$ perturbation parameters for the two cases are shown in Fig.~\ref{fig:correlation_quark-1}. The layout of the figure is the same as that of Fig.~\ref{fig:correlation-1}. The conclusions pertaining to Fig.~\ref{fig:correlation_quark-1} are also qualitatively similar to that of Fig.~\ref{fig:correlation-1}. While $0.03 < |d_u| <0.04$ is necessary for \emph{Case~I}, a broad range of values of $a_u$ and $b_u$ are allowed for \emph{Case~II}. No preference for the phase is evident in either case. 

For the down quark mass matrix, our starting point is again a democratic mass matrix since down quarks obtain their mass in an initially similar manner to charged leptons and up quarks. However, as discussed earlier, the residual $\mathbb{S}_2^L \times \mathbb{S}_2^R$ symmetry needs to be completely broken in order to obtain nonzero values of $\theta_{12}$ and $\theta_{13}$ in $V_{\text{CKM}}$.
Thus, the full down quark mass matrix is parametrized as
\beq
M_d = \dfrac{M_{0}}{3} \, \Bigg[ \begin{pmatrix}
1 & 1 & 1\\
1 & 1 & 1\\
1 & 1 & 1
\end{pmatrix} \, + \,   \begin{pmatrix}
\epsilon_{d_{11}} e^{i \alpha_{d_{11}}} & \epsilon_{d_{12}} e^{i \alpha_{d_{12}}} & \epsilon_{d_{13}} e^{i \alpha_{d_{13}}}\\
\epsilon_{d_{21}} e^{i \alpha_{d_{21}}} & \epsilon_{d_{22}} e^{i \alpha_{d_{22}}} & \epsilon_{d_{23}} e^{i \alpha_{d_{23}}}\\
\epsilon_{d_{31}} e^{i \alpha_{d_{31}}} & \epsilon_{d_{32}} e^{i \alpha_{d_{32}}} & \epsilon_{d_{33}} e^{i \alpha_{d_{33}}}
\end{pmatrix} \Bigg] \, ,
\eeq
where $\epsilon_{d_{ij}} \leq 0.1$ and $\alpha_{d_{ij}} \in [0, 2\pi]$. These parameters may be interpreted as a combination of $\mathbb{S}_2^L \times \mathbb{S}_2^R$ invariant and random perturbations.
We  diagonalize the down quark mass matrix and ensure that $m_s/m_b$ and $m_d/m_b$ lie within the $3 \sigma$ range of  $m_s(M_Z)/m_b(M_Z)$ and $m_d(M_Z)/m_b(M_Z)$ (evaluated from their measured values)~\cite{Running_masses}, and the $V_{\text{CKM}}$ parameters $\theta_{12},\theta_{13},\theta_{23}$, and $\delta_{\text{CP}}$~\cite{UTFIT} are also within $3 \sigma$ of their observed values. In Table~\ref{tab:quark-I} we list the perturbation parameters of benchmark points for each case that satisfy all the constraints. 
 

\begin{table}[t]
\resizebox{\columnwidth}{!}{%
\begin{tabular}{|c|c| c| c|c c c|c c c|c c c|c c c c| }
\hline
Case & \multicolumn{3}{|c|}{Up sector perturbations} & \multicolumn{6}{|c|}{Down sector perturbations}  & \multicolumn{3}{|c|}{Mass ratios} & \multicolumn{4}{|c|}{$V_{\text{CKM}}$ observables}
\\
\hline
 & $|d_u|$ or $a_u$ & $b_u$ & $\alpha_{I_u}$ or $\alpha_{II_{u}}$& \multicolumn{3}{|c|}{$\epsilon_{d_{ij}}$}  & \multicolumn{3}{|c|}{$\alpha_{d_{ij}}$} & $\dfrac{m_c}{m_t}$ & $\dfrac{m_d}{m_b}$ &  $\dfrac{m_s}{m_b}$ & $\theta_{12}$ & $\theta_{13}$ & $\theta_{23}$ & $\delta_{\text{CP}}$ \\
 & & & [Radian] & & & & \multicolumn{3}{|c|}{[Radian]} & & & & \multicolumn{4}{|c|}{$[\circ]$}\\
\hline
\hline 
\multirow{3}{*}{\emph{I}} & \multirow{3}{*}{0.015} & \multirow{3}{*}{-} & \multirow{3}{*}{4.84} & 0.016 & 0.062 & 0.032 & 5.44 & 4.98 & 4.76 & \multirow{3}{*}{0.00327} & \multirow{3}{*}{0.00152} & \multirow{3}{*}{0.02910} & \multirow{3}{*}{12.98} & \multirow{3}{*}{0.194}  & \multirow{3}{*}{2.427} & \multirow{3}{*}{60.5} \\

 &  &  &  & 0.008 & 0.072 & 0.027 & 1.49 & 4.68 & 5.03 & & & & & & &\\
 &  &  &  & 0.080 & 0.099 & 0.002 & 0.83 & 1.38 & 0.71 & & & & & & &\\
\hline
\multirow{3}{*}{\emph{II}} & \multirow{3}{*}{0.445} & \multirow{3}{*}{0.351} & \multirow{3}{*}{2.77} & 0.017 & 0.002 & 0.011 & 2.42 & 1.11 & 5.15 & \multirow{3}{*}{0.00274} & \multirow{3}{*}{0.00132} & \multirow{3}{*}{0.03047} & \multirow{3}{*}{12.89} & \multirow{3}{*}{0.204}  & \multirow{3}{*}{2.484} & \multirow{3}{*}{64.5} \\

 &  &  &  & 0.034 & 0.020 & 0.027 & 2.49 & 4.60 & 5.33 & & & & & & &\\
 &  &  &  & 0.019 & 0.083 & 0.074 & 4.85 & 2.39 & 2.88 & & & & & & &\\
\hline

\end{tabular}
}
\caption{ An illustrative set of perturbation parameters for the up and down quark sectors, and the corresponding values of quark sector mass and mixing observables for \emph{Cases I} and \emph{II}.   }
\label{tab:quark-I}
\end{table}


Having established that $\mathbb{S}_2^L \times \mathbb{S}_2^R$ perturbations in the charged lepton and up quark sectors along with arbitrary perturbations in the down quark and neutrino sectors provide a consistent model, it is clear that the same is true if the residual symmetry is  
$\mathbb{S}_2$ because of the larger number of independent parameters in the corresponding perturbation mass matrix.

\section{Conclusions}
\label{sec:conclusion}

We consider a broken democracy model to explain the mass and mixing patterns in both lepton and quark sectors. We begin with democratic mass matrices for the charged lepton, up quark and down quark sectors, and a diagonal mass matrix for the neutrino sector. The difference in the structure of neutrino mass matrix from the other sectors can be understood by a distinct breaking of $O(3)_L$ symmetry in the neutrino sector. Although the mass matrices outlined above naturally explain 
two large mixing angles in $V_{\text{PMNS}}$, vanishing masses for the first two generations for both charged leptons and quarks are predicted. Hence, to explain the observed fermion masses and mixings, the $\mathbb{S}_3^L \times \mathbb{S}_3^R$ symmetry in the democratic mass matrices needs to be broken.

 
We propose a model in which $\mathbb{S}_3^L \times \mathbb{S}_3^R$ is broken to a smaller discrete symmetry, $\mathbb{S}_2^L \times \mathbb{S}_2^R$, which is manifested as small perturbations to democratic mass matrices. 
$\mathbb{S}_2^L \times \mathbb{S}_2^R$ perturbations in the charged lepton sector alone can not explain the observed solar ($\theta_{12}$) and the reactor ($\theta_{13}$) mixing angles of neutrinos. We find a remedy to this problem by introducing arbitrary perturbations in the neutrino mass matrix, which can be generated by assuming that there is no remnant symmetry in the neutrino sector. We analyze two special cases of $\mathbb{S}_2^L \times \mathbb{S}_2^R$ invariant perturbations and find solutions that satisfy all mass and mixing observables in the lepton sector within $3 \sigma$ of their measured values.

On the other hand, for the quark sector, it is expected that the perturbation matrices in both the up and down sector should preserve an $\mathbb{S}_2^L \times \mathbb{S}_2^R$ symmetry since they obtain their masses in a similar fashion to charged leptons. However, this leads to $\theta_{12}=\theta_{13}=0$ in the CKM matrix. We therefore assume that the residual $\mathbb{S}_2^L \times \mathbb{S}_2^R$ symmetry in the down quark mass matrix is entirely broken and the resulting perturbations are arbitrary. Under these assumptions, we obtain solutions consistent with up-type and down-type quark mass ratios and the CKM parameters within $3 \sigma$ of their experimental values. 
    
  Note that the remnant $\mathbb{S}_2^L \times \mathbb{S}_2^R$ symmetries in the charged lepton and up quark mass matrices may also be broken. However, since the $\mathbb{S}_2^L \times \mathbb{S}_2^R$ invariant perturbations already provide a viable solution for these sectors, such a symmetry breaking has to be much softer than for down quarks. This is not unnatural since the hierarchy of down quark masses is relatively mild compared to that of charged leptons and up quarks. Our analysis shows that a consistent flavor model for SM fermions results from a breaking pattern, $\mathbb{S}_3^L \times \mathbb{S}_3^R \rightarrow \mathbb{S}_2^L \times \mathbb{S}_2^R$, with eventual breaking of the residual $\mathbb{S}_2^L \times \mathbb{S}_2^R$ symmetry. 
  One can also construct models by replacing $\mathbb{S}_2^L \times \mathbb{S}_2^R$ by the smaller symmetry $\mathbb{S}_2$.        

\begin{acknowledgments}
This work was supported by U.S. DOE Grant No.~de-sc0010504, U.S. NSF Grant No.~PHY-1250573, and by KAKENHI, Japan, Grant Nos. 26104001, 26104009, 16H02176, and 17H02878. T.T.Y. is also supported by a Hamamatsu Professorship.
\end{acknowledgments}



\end{document}